\documentclass[english,twocolumn,prb,showpacs]{revtex4}
\usepackage[]{fontenc}
\usepackage[latin1]{inputenc}
\usepackage{amsmath}
\usepackage{graphicx}
\usepackage{amssymb}
\usepackage{psfrag}

\newcommand{\bvec}[1]{{\mathbf #1}}

\usepackage{babel}

\begin{document}

\title{Magnetotransport in disordered delta-doped heterostructures}

\author{V. Tripathi$^{1,2}$ and M. P. Kennett$^{3}$}

\affiliation{$^{1}$ Theory of Condensed Matter Group, Cavendish Laboratory, Department
of Physics, University of Cambridge, J. J. Thomson Avenue, Cambridge
CB3 0HE, United Kingdom}
\affiliation{$^{2}$ Department of Theoretical Physics, Tata Institute of Fundamental Research, Homi Bhabha oad, Mumbai 400005, India}
\affiliation{$^{3}$ Physics Department, Simon Fraser University, 8888 University
Drive, Burnaby, BC, V5A 1S6, Canada}

\date{\today}

\begin{abstract}
We discuss theoretically how electrons confined to two dimensions
in a delta-doped heterostructure can arrange themselves in a droplet-like
spatial distribution due to disorder and screening effects when their
density is low. We apply this droplet picture to magnetotransport
and derive the expected dependence on electron density of several
quantities relevant to this transport, in the regimes of weak and
moderate magnetic fields. We find good qualitative and quantitative
agreement between our calculations and recent experiments on delta-doped
heterostructures. 
\end{abstract}


\maketitle

\section{Introduction}

Delta doping of semiconductor heterostructures has a history of leading
to new and interesting physical phenomena, such as the fractional
quantum Hall Effect. \cite{Tsui} This new physics has come from increasing
the mobility of doped electrons by minimizing the disorder they feel
due to the spatial separation between the dopants and the electrons.
While spatial separation reduces the magnitude of fluctuations in
the potential due to dopant atoms, it is well known that disorder
due to the random positions of donors in the delta-doped layer has
important consequences for the transport properties of heterostructures.
Recent efforts have attempted to gain a clearer experimental picture
of the nano-scale distribution of electronic states in two dimensional
electron gases (2DEGs).\cite{Yacoby1,Yacoby2,Ilani2,Wiebe} Such electronic
inhomogeneity on the nano-scale is believed to be important for transport
in a wide variety of strongly correlated electronic materials.\cite{Davis,Various,Manganites}

A step towards a more systematic understanding of the effect of disorder
on transport properties has been taken in recent experiments on delta-doped
devices by Baenninger \textit{et al.} \cite{Baenninger1,Baenninger2}
where a series of small devices was fabricated in which the distance
between the dopant layer and the 2DEG was varied in a controlled way.
Some of these devices show very interesting resistance effects in
a magnetic field. In particular, the dependence of the magnetoresistance
on electron density has been suggested as evidence for charge density
wave formation.\cite{Baenninger2}

In this paper, we argue that the observations of Ghosh \textit{et
al.} \cite{Baenninger2,Ghosh3,Ghosh2,Ghosh1} are a manifestation
of charge droplet formation in a 2DEG. To justify this point of view,
we analyze the charge distribution in a disordered 2DEG due to a delta
doped layer, to demonstrate that in the devices of interest, the charge
distribution likely consists of droplets of charge that sit in minima
of the screened disorder potential. It is well known theoretically
that the charge distribution in delta-doped heterostructures should
have a droplet-like structure at low electron density.\cite{Suris,Nixon}
There has also been recent interest in electron droplets in the vicinity
of the 2D metal insulator transition,\cite{Shi,Fogler} but relatively
little attention has been paid to transport in insulating samples
in a magnetic field (with the exception of Efros and co-workers\cite{Efros1,Efros2,EPB}).

We derive expressions for the physical parameters of electron droplets
in the non-linear screening regime that is relevant to the experiments
of interest here, and then apply this picture to describe the transport
in these devices. We observe that our picture implies that the tunneling
between droplets decreases as a function of magnetic field in a manner
that is consistent with experiment. The picture for the magnetoresistance
at small fields is along the lines of that proposed by Glazman and
Raikh,\cite{Glazman} while at larger fields, the magnetoresistance
crosses over to the behavior expected by Shklovskii and Efros.\cite{Shklovskii1,Shklovskii2}
These approaches\cite{Glazman,Shklovskii1,Shklovskii2} provide the
magnetic field dependence of the resistance, but the picture of electron
droplets yields non-trivial predictions for the dependence of resistance
on parameters other than magnetic field, such as electron and dopant
densities. In particular we show that in the regime of magnetic fields
where the resistivity $\rho$ varies with magnetic field $B$ as $\rho(B)\propto\exp[\alpha B^{2}]$,
that $\alpha\propto n_{e}^{-\frac{3}{2}}$, where $n_{e}$ is the
electron density in the 2DEG, even though the average tunneling distance
between droplets is much larger than the average inter-electron spacing.
We also expect $\alpha$ to be temperature-independent at low temperatures,
as observed in experiment.

This paper is structured as follows: in Sec.~\ref{sec:model} we
introduce our model for the delta doped heterostructure and elucidate
the mechanism by which electron droplets form, then characterize how
their properties depend on sample parameters. In Sec.~\ref{sec:magnetotransport}
we discuss the expectations for magnetotransport, based on the droplet
picture, and compare the implications with actual transport measurements.
Finally, in Sec.~\ref{sec:conc} we conclude and discuss the implications
of our results for experiment.

\section{Model}

\label{sec:model}

In this section we introduce the physical situation that we are interested
in, and the model we have of this system. We then derive in detail
how disorder and screening effects give rise to a droplet-like electron
density distribution in the 2DEG, and give the physical parameters
of the electron droplets. We finally present a brief summary of the
electron droplet picture.

\subsection{Main parameters}

\begin{figure}
\includegraphics[width=7cm,keepaspectratio]{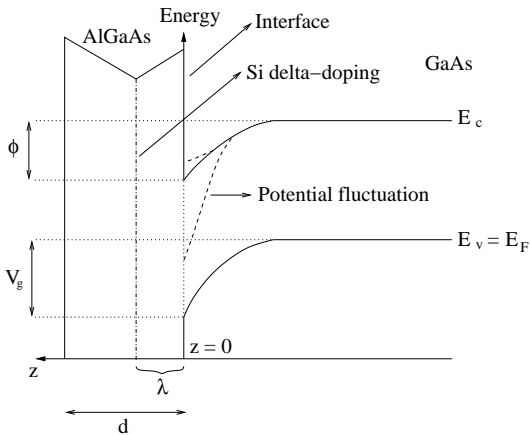}
\caption{Bandstructure of the heterostructure indicated by bold solid lines.
The 2DEG is formed at the interface $z=0$ of AlGaAs and GaAs. The
dashed lines indicate the random positions of the bottom of the conduction
band in the 2DEG; the randomness arises from a spatially nonuniform
distribution of $\delta-$dopants. The potential $\phi$ at the interface
is due to the gate voltage $V_{g},$ the ionized dopants' potential
and screening by conduction electrons in the interface. $E_{c}$ and
$E_{v}$ are the conduction and valence band energies, respectively,
in bulk GaAs and $E_{F}$ is the Fermi energy. }
\label{fig:bands} 
\end{figure}

The physical situation and several of the relevant parameters for
the physics we will consider in more detail later are summarized in
Fig.~\ref{fig:bands}.

There are two classes of parameters that control the physics in the
devices of interest: those that are intrinsic to GaAs, and those that
can be varied experimentally, through fabrication of a device, or
use of a gate. The properties that are intrinsic to GaAs are $\kappa=12.9,$
the dielectric constant of the GaAs or AlGaAs dielectric, and $m=0.067m_{e},$
the effective electron mass in GaAs, which imply a Bohr radius for
electrons in the 2DEG of $a_{B}=4\pi\kappa\epsilon_{0}\hbar^{2}/me^{2}\simeq10\,{\rm nm}.$
The following properties can be tuned in experiments:

\begin{enumerate}
\item $d,$ the distance between the gate electrode and the 2DEG. Experimentally,
$d\approx300\,{\rm nm}.$ 
\item $\lambda,$ the distance between the $\delta-$doping layer from the
2DEG. $\lambda=10-80\,{\rm nm}.$ The typical $\lambda$ used in our
calculations is $50$ nm. 
\item $n_{d}\sim10^{12}\,{\rm cm}^{-2},$ the density of dopants in the
$\delta$ layer. 
\item $n_{e}\sim10^{11}\,{\rm cm}^{-2},$ the average density of electrons
in the 2DEG in a typical sample. This will be the typical value in
our calculations. Some samples have lower electron densities, $n_{e}\sim10^{10}\,{\rm cm}^{-2}.$ 
\item $V_{g},$ the voltage at the gate electrode, which is metallic. 
\end{enumerate}
In our theoretical treatment, we always assume that $d\gg\lambda,$
and $n_{d}\gg n_{e}.$ The dielectric constants for GaAs and AlGaAs
are also assumed to be the same without loss of generality. 

\subsection{Electron droplet formation in a 2DEG}

\label{sec:droplet}

Our discussion follows similar lines to the work of Gergel' and Suris,
\cite{Suris} who considered the formation of electron droplets in
a 2DEG when the average position of the bottom of the conduction band
(see Fig.~\ref{fig:bands}) is above the Fermi energy $E_{F}.$ However,
our results differ qualitatively from Ref.~\onlinecite{Suris},
and these differences are essential for connecting our results with
experiment. Hence we present the effect of various experimentally
relevant parameters on the properties of the droplets in some detail.

We assume that the donor positions are uncorrelated, and distributed
with a white-noise Gaussian distribution, and that the donors are
ionized after compensating the band-bending field, which provides
the electrons for the two dimensional electron gas. We ignore effects
due to incomplete and weak ionization. \cite{Pikus} For a mean $\delta$-doping
density of $n_{d}=10^{12}{\,{\rm cm}}^{-2}$ and a 2DEG-dopant layer
separation of $\lambda=50{\,{\rm nm}},$ a complete ionization of
the donors corresponds to a compensating potential $V_{{\rm comp}}=e\lambda n_{d}/(\kappa\epsilon_{0})\sim1\,{\rm V}.$
Random fluctuations in the dopant potential cause the bottom of the
conduction band in some regions of the 2DEG to lie below $E_{F};$
this leads to the appearance of electron droplets. In the rest of
our discussion, we consider the potential $\phi$ in the 2DEG after
subtracting the compensating potential. Charge fluctuations about
the mean can be of either sign. We also assume that the chemical potential
is uniform across the sample.

The charge density $\rho(z,\mathbf{r})$ in the AlGaAs region is\begin{align}
\rho(z,\mathbf{r}) & =en(\mathbf{r})\delta(z-\lambda),\label{eq:rho}\end{align}
 where $n(\mathbf{r})=n_{+}(\mathbf{r})-n_{-}(\mathbf{r})$ is the
surface charge density in the $\delta$-layer and $\mathbf{r}$ is
a two-dimensional vector in the $\delta$-doping plane. The subscripts
$\pm$ denote the sign of the charge fluctuation. The spatial distribution
of the dopant atoms satisfies \begin{align}
\langle n(\mathbf{r})n(\mathbf{r}')\rangle-\langle n\rangle^{2} & =n_{d}\delta(\mathbf{r}-\mathbf{r}'),\label{eq:gaussian}\end{align}
 and $n_{d}=\langle n_{+}\rangle+\langle n_{-}\rangle$ is the total
surface charge density of \emph{both} polarities in the $\delta-$layer.
The dopant atoms create a fluctuating potential $\phi(\mathbf{r})$
in the 2DEG,

\begin{align}
\phi(\mathbf{r}) & =V_{{\rm comp}}+\frac{1}{4\pi\epsilon_{0}\kappa}\int\!\! d\mathbf{r}'\!\!\int_{0}^{d}\!\!\!\! dz\,\left\{ \rho(z,\mathbf{r}')-en_{e}\left[\phi(\mathbf{r}')\right]\delta(z)\right\} \nonumber \\
\times & \left[\frac{1}{\sqrt{(\mathbf{r}-\mathbf{r}')^{2}+z^{2}}}-\frac{1}{\sqrt{(\mathbf{r}-\mathbf{r}')^{2}+(2d-z)^{2}}}\right].\label{eq:potential1}\end{align}
 In Eq.~(\ref{eq:potential1}), $n_{e}[\phi(\mathbf{r}')]$ is the
electron charge density in the 2DEG and the first and second terms
under the square root signs represent direct and image contributions
from the charge distribution respectively. The position of the image
charges is to the left of the metallic gate electrode $z=d$ in Fig.~\ref{fig:bands}.

When the gate voltage $V_{g}$ is small, the number of electrons $n_{e}$
at the interface will be small. Let us first estimate the mean square
value of the potential fluctuations, $\langle\delta\phi^{2}\rangle,$
at the interface by ignoring screening effects of finite electron
density, $n_{e}$. Using Eqs.~(\ref{eq:gaussian}) and (\ref{eq:potential1})
one has \begin{align}
\langle\delta\phi^{2}\rangle & =\frac{2\pi n_{d}e^{2}}{(4\pi\kappa\epsilon_{0})^{2}}\int dr\, r\,\left[\frac{1}{\sqrt{r^{2}+\lambda^{2}}}\right.\nonumber \\
 & \left.\qquad\qquad\qquad\qquad\qquad-\frac{1}{\sqrt{r^{2}+(2d-\lambda)^{2}}}\right]^{2}\nonumber \\
 & =\frac{n_{d}e^{2}}{8\pi\kappa^{2}\epsilon_{0}^{2}}\ln\left[\frac{4d^{2}}{\lambda(2d-\lambda)}\right].\label{eq:meansqpoten}\end{align}
 When the gate voltage $V_{g}$ is increased, the bottom of the conduction
band approaches the Fermi energy $E_{F}.$ If potential fluctuations
due to the dopants are such that in some regions of the interface,
the bottom of the conduction band now lies below $E_{F},$ a finite
$n_{e}(\mathbf{r})$ will be found at such places (see Fig.~\ref{fig:bands}).

Next we estimate the effect of screening on the mean square fluctuation
$\langle\delta\phi^{2}\rangle$ due to a local electron density $n_{e}(\mathbf{r})$.
Consider a region of size $R$ in the $\delta$-layer. From Eq.~(\ref{eq:gaussian}),
one can estimate the magnitude of fluctuations of $n(\mathbf{r})$
in this region:

\begin{align}
\left<\delta n^{2}(R)\right> & =\frac{1}{\pi R^{2}}\int_{0}^{R}d\mathbf{r}'(\langle n(\mathbf{r})n(\mathbf{r}')\rangle-\langle n\rangle^{2})=\frac{n_{d}}{\pi R^{2}}.\label{eq:fluctuation1}\end{align}
 Thus $\sqrt{\left<\delta n^{2}(R)\right>}\sim n_{d}^{1/2}/\pi^{1/2}R$
is the characteristic fluctuation of the dopant charge density. If
the local electron charge density, $n_{e}(\bvec{r}),$ exceeds $\sqrt{\left<\delta n^{2}(R)\right>},$
then potential fluctuations due to these fluctuations in charge density
will be screened by a redistribution of the electronic charge $n_{e}.$
If the local electron charge density is much less than $\sqrt{\left<\delta n^{2}(R)\right>},$
then potential fluctuations in a region of size $R$ are not screened.
Therefore, for a given electron density $n_{e},$ we only need to
take into account fluctuations in regions of size $R<R_{c}=n_{d}^{1/2}/\pi^{1/2}n_{e}.$
Thus the upper limit of integration over $r$ in Eq.~(\ref{eq:meansqpoten})
can be replaced with $R_{c}=n_{d}^{1/2}/\pi^{1/2}n_{e};$ consequently

\begin{align}
\langle\delta\phi^{2}\rangle & =\frac{n_{d}e^{2}}{8\pi\kappa^{2}\epsilon_{0}^{2}}\left\{ \ln\left[\frac{4d^{2}}{\lambda(2d-\lambda)}\right]\right.\nonumber \\
- & \left.2\ln\left[\left(\frac{\lambda^{2}+R_{c}^{2}}{(2d-\lambda)^{2}+R_{c}^{2}}\right)^{1/4}\!\!\!\!\!+\left(\frac{(2d-\lambda)^{2}+R_{c}^{2}}{\lambda^{2}+R_{c}^{2}}\right)^{1/4}\right]\right\} .\label{eq:meansqpoten2}\end{align}
 Equation~(\ref{eq:meansqpoten2}) yields two simple limiting regimes
for the relevance of screening effects:

\begin{align}
\langle\delta\phi^{2}\rangle & =\frac{n_{d}e^{2}}{8\pi\kappa^{2}\epsilon_{0}^{2}}\times\left\{ \begin{array}{lr}
\ln\left(\frac{2d}{\lambda}\right), & R_{c}\gg2d\gg\lambda\\
\frac{1}{2}\ln\left(1+\left(\frac{R_{c}}{\lambda}\right)^{2}\right), & 2d\gg R_{c},\lambda\end{array}\right..\label{eq:meansqpotlimits}\end{align}
 The physical picture behind Eq.~(\ref{eq:meansqpotlimits}) is as
follows. The metal electrode screens potential fluctuations on a length
scale of $d.$ Thus, if the fluctuations of the dopant density give
a screening radius $R_{c}$ that is larger than $d,$ we may ignore
the screening effects of a finite electron density $n_{e}.$ For the
devices we are interested in, the important limit is generally where
$d$ is considerably larger than $\lambda$, although $R_{c}$ may
be of the same order as, or greater than $d$ at very low electron
densities. We also note that $R_{c}$ cannot fall below $n_{d}^{-1/2}$,
since the Gaussian approximation {[}Eq.~(\ref{eq:gaussian})] that
we used to determine it will no longer be valid.

Let us now estimate the spatial dimensions of the electron droplets
localized in the minima of the smoothly fluctuating 2DEG potential.
We first do this for moderate electron densities, for which we can
assume $2d\gg R_{c}$, to illustrate our logic, and then also consider
the situation $2d\sim R_{c}$, where one has to deal with 
Eq.~(\ref{eq:meansqpoten2}) numerically.

From Eq.~(\ref{eq:meansqpotlimits}), it is clear that the confining
potential increases as the distance to the $\delta$-layer, $\lambda,$
is decreased. Nevertheless, the kinetic energy cost of electron confinement
in the $z$-direction means that the localization length in the $z$-direction
does not vanish. We assume that the electron wavefunction is localized
in the $z$-direction in the GaAs substrate with a characteristic
lengthscale $z_{0}$. Hence, the kinetic energy of the electron is
of the order of $n^{2}\hbar^{2}\pi^{2}/(2mz_{0}^{2}),$ ($n$ is a
natural number) and the typical distance of the electron from the
dopant layer is of the order of $\lambda+z_{0}.$ The potential fluctuation
expression in Eq.~(\ref{eq:meansqpotlimits}) then needs to be modified:
\begin{align}
e\sqrt{\langle\delta\phi^{2}\rangle} & =\frac{e^{2}n_{d}^{1/2}}{4\sqrt{\pi}\kappa\epsilon_{0}}\left\{ \ln\left[1+\left(\frac{R_{c}}{\lambda+z_{0}}\right)^{2}\right]\right\} ^{\frac{1}{2}}.\label{eq:rmspotential}\end{align}
 We now minimize the total energy with respect to $z_{0,n}$ to get
the binding energy $E_{n}$ for the $n^{{\rm th}}$ sub-band,\cite{Suris}
where

\begin{align}
E_{n} & =\frac{\hbar^{2}\pi^{2}n^{2}}{2mz_{0,n}^{2}}-\frac{n_{d}^{1/2}e^{2}}{4\sqrt{\pi}\kappa\epsilon_{0}}\left\{ \ln\left[1+\left(\frac{R_{c}}{\lambda+z_{0,n}}\right)^{2}\right]\right\} ^{\frac{1}{2}}.\label{eq:Ez0}\end{align}
 A bound state is always possible with such a potential because while
the kinetic energy decreases as $1/z_{0,n}^{2},$ the magnitude of
the potential energy decreases only as $1/z_{0,n}.$ The minimization
leads to the following transcendental equation for $z_{0,n}$:

\begin{eqnarray}
z_{0,n}^{3} & = & \frac{n^{2}\pi^{\frac{3}{2}}a_{B}}{n_{d}^{\frac{1}{2}}}(\lambda+z_{0,n})\left[1+\left(\frac{\lambda+z_{0,n}}{R_{c}}\right)^{2}\right]\nonumber \\
 &  & \qquad\times\left\{ \ln\left(1+\left(\frac{R_{c}}{\lambda+z_{0,n}}\right)^{2}\right)\right\} ^{\frac{1}{2}}.\label{eq:z0_n}\end{eqnarray}
 Eq.~(\ref{eq:z0_n}) gives good results when $2d\gg R_{c}.$ For
values of the electron density where this condition is not well-satisfied,
then in Eq.~(\ref{eq:Ez0}), the more accurate expression for potential
fluctuations, Eq.~(\ref{eq:meansqpoten2}) should be used. Table~\ref{tab:z0}
shows the numerically calculated values of $E_{1},$ $E_{2},$ $z_{0,1}$
and $z_{0,2}$ using Eq.~(\ref{eq:meansqpoten2}), with $\lambda$
replaced by $\lambda+z_{0},$ for the potential fluctuations.

\begin{table}
\begin{tabular}{|c|c|c|c|c|c|c|c|}
\hline 
$n_{e}$(cm$^{-2}$)&
$R_{c}$&
$E_{1}$(K)&
$E_{2}$(K)&
$z_{0,1}$&
$R_{p}$&
$\Delta$(K)&
$\xi$\tabularnewline
\hline
\hline 
$10^{11}$&
56&
-69&
-29&
49&
52&
45&
17\tabularnewline
\hline 
$5\times10^{10}$&
113&
-133&
-77&
43&
50&
47&
9\tabularnewline
\hline 
$2\times10^{10}$&
282&
-202&
-141&
42&
50&
47&
6.5\tabularnewline
\hline
$10^{10}$&
564&
-221&
-159&
42&
50&
47&
6\tabularnewline
\hline
\end{tabular}\bigskip{}
\caption{Values of the screening radius $R_{c},$ the bound state energies
$E_{1}$ and $E_{2},$ and the penetration distance for the lowest
sub-band $z_{0,1}$ in the GaAs layer for different values of electron
density $n_{e}.$ Also shown are the droplet sizes $R_{p},$ the highest
energy $\Delta$ of the electrons occupying a droplet, and the localization
lengths $\xi$ for inter-droplet tunneling. The lengths $R_{c},$
$z_{0,1},$ $R_{p}$ and $\xi$ are all in units of nanometers.}
\label{tab:z0}
\end{table}

The 2DEG electrons fill the lowest sub-band first and the second sub-band
does not begin to be populated until the most energetic electron in
the lower level reaches $E_{2}.$ The basic result of this analysis
is that droplets can form due to electrons filling the minima of the
screened disorder potential. 

We now consider the dimension of the droplets in the plane of the
2DEG. The following arguments are valid for both $z_{0}>\lambda$
and $z_{0}<\lambda.$ In a region of size $R\ll R_{c},$ the charge
density fluctuation is of the order of $n_{d}^{1/2}/\pi^{1/2}R\gg n_{e}.$
The electrons begin filling these small but deep regions first in
an attempt to screen the strongest charge fluctuations and the density
$n_{e,\text{local}}=n_{d}^{1/2}/\pi^{1/2}R$ of electrons in these
droplets will be much larger than the mean 2DEG electron density $n_{e}.$
However, since the kinetic energy rises with confinement, this limits
the density from becoming too large, since to high a density would
prevent the formation of bound states. With an electron density of
$n_{e,\text{local}}$, correlations of the potential beyond the length
scale $R$ will get screened. Using the arguments we employed for
$\langle\delta\phi^{2}\rangle$ earlier, we can similarly say that
the potential fluctuation with an electron density of $n_{e,\text{local}}$
will be of the order of \begin{equation}
\left(\frac{e^{2}n_{d}^{1/2}}{4\sqrt{\pi}\kappa\epsilon_{0}}\right)\left\{ \ln\left[1+\left(\frac{R}{\lambda+z_{0}}\right)^{2}\right]\right\} ^{\frac{1}{2}}.\label{eq:localfluc}\end{equation}
 In an electron droplet of size $R_{p}$, the kinetic energy of the
electron occupying the highest state in the droplet will be of the
same order as the potential energy fluctuation. If $R_{p}$ is small
compared to $\lambda+z_{0}$ (we shall see shortly that this is the
case), the local fluctuation, Eq.~(\ref{eq:localfluc}) is a linear-$R$
confinement. For such a confinement, the virial theorem suggests that
the kinetic energy should be half of the potential energy. For the
lowest sub-band, $n=1,$ we have

\begin{align}
\frac{\hbar^{2}k_{\text{max}}^{2}}{2m} & =\frac{1}{2}\times\frac{e^{2}n_{d}^{1/2}}{4\sqrt{\pi}\kappa\epsilon_{0}}\left[\ln\left(1+\left(\frac{R_{p}}{\lambda+z_{0,1}}\right)^{2}\right)\right]^{\frac{1}{2}}.\label{eq:dropletsize1}\end{align}
 It should be noted that we can only make an order of magnitude comparison,
as obtaining the exact magnitude of the screened Coulomb fluctuations
where the electrons are pooled is beyond the scope of our discussion.
To estimate the wavevector $k_{\text{max}}$ we note that the droplet
is effectively two-dimensional since motion deep into the GaAs substrate
is not possible. Equating the number of droplet eigenstates, $(k_{\text{max}}R_{p})^{2}/2,$
with the charge fluctuation $N_{e}$ in the droplet, $N_{e}=\pi^{1/2}n_{d}^{1/2}R_{p},$
(the number of electrons screening the potential fluctuation on a
length scale $R$), we have $k_{\text{max}}^{2}=2n_{d}^{1/2}\pi^{1/2}/R_{p}.$
Using this estimate for $k_{\text{max}}$ in Eq.~(\ref{eq:dropletsize1}),
we arrive at the condition

\begin{align}
\frac{a_{B}}{R_{p}} & =\frac{1}{2}\left[\ln\left(1+\left(\frac{R_{p}}{\lambda+z_{0,1}}\right)^{2}\right)\right]^{\frac{1}{2}},\label{eq:dropletcondition}\end{align}
 and when we specialize to the case $\lambda+z_{0,1}\gg a_{B},$ the
droplet condition, Eq.~(\ref{eq:dropletcondition}) gives \begin{align}
R_{p} & \sim R_{p,1}=\sqrt{2a_{B}(\lambda+z_{0,1})},\label{eq:dropletsize2}\end{align}
 as the size of the droplet in the lowest sub-band. We note that in
Ref.~\onlinecite{Suris} the form of the potential energy used
to estimate $R_{p}$ was for the $\lambda\ll R$ limit, however, the
conclusion of the analysis was that $R_{p}\sim\lambda$. This invalidates
the original assumption and Eq.~(\ref{eq:dropletcondition}) should
be used instead. Equation~(\ref{eq:dropletsize2}) gives for $n_{e}=5\times10^{10}\,{\rm cm}^{-2}\text{ and }z_{0,1}\simeq43\,{\rm nm},$
$R_{p,1}\simeq43\,{\rm nm}.$ This compares well with the numerical
estimate, $R_{p}\simeq50\,{\rm nm}$ (see Table \ref{tab:z0}). 

The estimate for $R_{p}$ can be further improved by noting that $z_{0}$
itself varies with $R_{p}.$ Physically, the confining potential is
weaker at smaller values of $R_{p},$ therefore $z_{0}$ increases
as $R_{p}$ is decreased. This effect will lead to an enhancement
of $R_{p}.$ For this purpose, one needs to minimize Eq.~(\ref{eq:Ez0})
for $z_{0}$ as a function of $R_{p}\ll(\lambda+z_{0}),$ and then
use the $R_{p}-$dependent $z_{0}$ in Eq.~(\ref{eq:dropletcondition})
to obtain the droplet size. We have not followed this quantitatively
more accurate but tedious procedure. Many of our quantitative estimates
will be affected by our inability, in the present paper, to get a
better estimate for $R_{p}.$ However our qualitative predictions,
and in particular, the dependence of quantities on experimental parameters
is not affected. In our quantitative estimates we will, for the sake
of greater accuracy, use the numerically computed values rather than
the analytic expressions that are strictly valid for $2d\gg R_{c}.$ 

The density of electrons in the droplets, $n_{e,\text{local}},$ is
a little greater than the average electron density, $n_{e}.$ For
example, for $n_{e}=5\times10^{10}\,{\rm cm}^{-2},$ we have $n_{e,\text{local}}=n_{d}^{1/2}/\pi^{1/2}R_{p}\simeq1.1\times10^{11}{\rm cm}^{-2}.$
The local variation in the electron density is due to the movement
of charge in the potential landscape from the {}``hills'' to the
{}``lakes''. The number of electrons in a droplet, $N_{e}=\pi^{1/2}n_{d}^{1/2}R_{p},$
is only weakly dependent on the average electron density through $z_{0}.$
Extra electrons are accommodated in the 2DEG through increasing the
density of droplets. The separation between the droplet centers,

\begin{align}
l_{ip} & =2(n_{d}^{1/2}R_{p}/\pi^{1/2}n_{e})^{1/2}=2\sqrt{R_{c}R_{p}},\label{eq:interdropletsep}\end{align}
decreases with increasing average electron density $n_{e}.$ We note
that a previous discussion of droplet formation in 2DEGs\cite{Suris}
states that $l_{ip}=R_{c}$ for the purposes of calculating the conductivity.
We correct that statement here, as our analysis makes clear that this
is not the case for the regimes of interest.

Next we discuss the height of the barrier separating two droplets,
$E_{\text{barrier}}$. This barrier is not simply the size of the
fluctuation from the bottom of the potential well, $e\sqrt{\langle\delta\phi^{2}\rangle}$,
which is what one would find for {}``empty'' potential wells. Instead
we need to consider the occupation of the wells, and hence $E_{\text{barrier}}$
is given by the difference between the magnitude of the binding energy
$E_{n}$ for confinement in the $z$ direction and $\Delta=\hbar^{2}k_{\text{max}}^{2}/2m=\hbar^{2}\pi^{1/2}n_{d}^{1/2}/(mR_{p}),$
the highest energy for the electrons occupying a well, reckoned from
the bottom of the well. The inter-droplet barrier height $E_{\text{barrier}}$
is hence given by $E_{\text{barrier}}=|E_{1}|-\Delta.$ Additionally,
the typical number of electrons in a droplet is $N_{e}=\pi^{1/2}n_{d}^{1/2}R_{p}\simeq9$
for $n_{e}=5\times10^{10}\,{\rm cm}^{-2}$ and $R_{p}=50\,{\rm nm}.$
Thus the mean level spacing in the droplets, $\delta\sim\Delta/N_{e}=\hbar^{2}/(mR_{p}^{2})\simeq5\,{\rm K}.$
As $\Delta$ is typically of the order of the sub-band spacing $|E_{1}-E_{2}|,$
the second sub-band in the $z$ direction is also likely to be populated
to a certain degree, although at low enough electron densities and
temperatures, this should not be a major concern, and due to the uncertainty
of our estimates, we shall assume that only the lowest sub-band is
populated. We do not expect this to significantly alter our conclusions.
The distance $r$ between the surfaces of two neighboring droplets
is \begin{align}
r & =l_{ip}-2R_{p}=2(\sqrt{R_{c}R_{p}}-R_{p}).\label{eq:r}\end{align}
 The localization length for inter-droplet tunneling can be obtained
from the size of the barrier, \begin{align}
\xi & =\hbar/\sqrt{2mE_{\text{barrier }}}=\hbar/\sqrt{2m(|E_{1}|-\Delta)}.\label{eq:xiloc}\end{align}
Table \ref{tab:z0} lists the calculated values of $\Delta$ and $\xi$
for different electron densities. Note that the localization length
here is of the order of the Bohr radius in GaAs. This should be regarded
as a coincidence. 

\subsection{Summary}

\begin{figure}
\psfragscanon \psfrag{catchment}{$\sqrt{R_{c}R_{p}}$} \psfrag{dropletsize}{$R_{p}$}
\psfrag{phi}{$\phi(r)$} \psfrag{2l}{$2R_{p}$} \psfrag{fluc}{$\sqrt{\langle\delta\phi^{2}\rangle}$}
\psfrag{D}{$\Delta$} \psfrag{r}{$r$} \includegraphics[width=7cm,keepaspectratio]{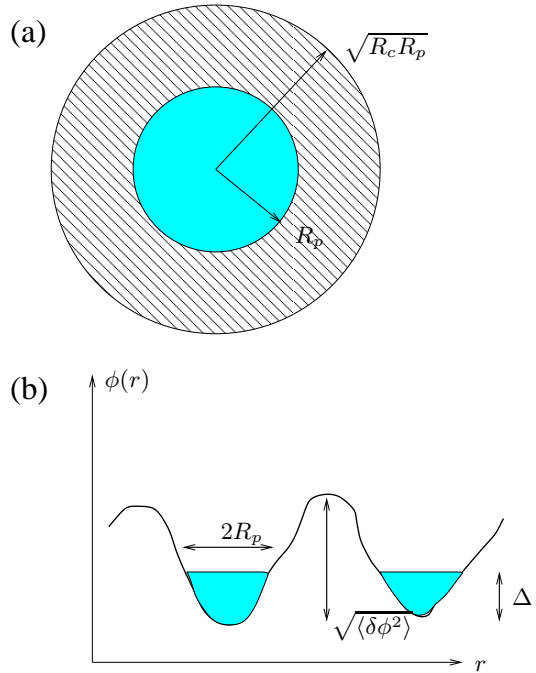}

\caption{Schematic picture of droplets. (a) A typical droplet (shown in solid
shading) has a radius of the order of the distance $\lambda$ between
the 2DEG and the $\delta-$layer. For a dopant density $n_{d},$ the
total number of electrons in the droplet is $N_{e}\sim n_{d}^{1/2}R_{p}.$
For an average 2DEG electron density $n_{e},$ the {}``catchment
area'' (shown in line shading) is $\sim n_{d}^{1/2}R_{p}/n_{e}=R_{c}R_{p}.$
Thus the inter-droplet distance is of the order of $\sqrt{R_{c}R_{p}}.$
(b) Cross-section of energy profile between droplets. The electrons
fill up to an energy $\Delta$ from the bottom of the potential wells.
The magnitude of the potential fluctuations is $|E_{1}|\sim e\sqrt{\langle\delta\phi^{2}\rangle}\gg\Delta.$}

\label{fig:droplets} 
\end{figure}

In the previous section (Sec.~\ref{sec:droplet}), we showed that
for low electron density delta doped heterostructures that give rise
to 2DEGs, it is very natural to expect that electrons will organize
themselves into droplets of charge, centered on the minima of the
screened disorder potential, for a quite reasonable range of parameters.
The important length and energy scales of the droplets that relate
to the transport properties of the device are the inter-droplet spacing,
and the energy barriers between droplets, as summarized in Fig.~\ref{fig:droplets}.

\section{Magnetotransport}

\label{sec:magnetotransport}

In this section we consider several aspects of magneto-transport.
In particular, we first consider the regime of low magnetic fields,
where, due to quantum interference effects, one generically expects
negative magnetoresistance, and we then move to higher magnetic fields,
where one expects positive magnetoresistance due to shrinking of the
localization length with increasing magnetic field. Our results are
in good agreement with experiment.

\subsection{Small fields: negative magnetoresistance}

\label{sub:negative magnres}

At small magnetic fields, many 2DEGs show negative magnetoresistance,
i.e. the resistance decreases with increasing magnetic field, which
is generically due to the magnetic field suppressing quantum interference
of different electron paths. The samples in Ref.~\onlinecite{Baenninger2}
show quadratic negative magnetoresistance at small fields: \begin{align}
[{\cal R}(B)-{\cal R}(0)]/{\cal R}(0) & =-(B/B_{c})^{2},\label{eq:negmagnres}\end{align}
 with $B_{c}\simeq0.15$ T. In the same magnetic field range, the
temperature dependence of the resistance obeys an Arrhenius law for
temperatures greater than about 1 K.

There are two contexts in which negative magnetoresistance in insulators
has been studied in the literature. The first is hopping conductivity
in dirty semiconductors, and the second is tunneling between droplets
of charge, similarly to the picture outlined in the previous section.

In dirty semiconductors, electrical conductivity occurs due to hopping
between different impurity sites and negative magnetoresistance is
a consequence of suppression of the destructive interference of the
Aharonov-Bohm phases acquired different trajectories.\cite{nguyen1,sivan,schirmacher}
Such a theory gives a negative magnetoresistance $\propto|B|$, the
perpendicular magnetic field, for fields such that $(BD^{3/2}\xi^{1/2})/\phi_{0}>e^{-D/\xi}.$
\cite{sivan,schirmacher} {[}$\xi$ is the localization length, $\phi_{0}=h/e$
is the flux quantum, and $D$ is the hopping distance.] For smaller
magnetic fields the negative magnetoresistance is quadratic in $B$.
If we use the value of $\xi$ and $D=r$ from our droplets scenario,
this criterion suggests the negative magnetoresistance will be quadratic
in the applied field up to $B\sim0.2\,$ T for $n_{e}=5\times10^{10}\,{\rm cm}^{-2}.$

In our case, the physical situation is somewhat different because
instead of hopping between impurity sites as in a dirty semiconductor,
the electrons in the 2DEG hop between droplets. The distance between
the centers of neighboring droplets, $l_{ip}=2\sqrt{R_{c}R_{p}},$
is comparable with the droplet diameter $2R_{p}$, unlike a dirty
semiconductor where the electrons are localized at point-like impurity
sites. This situation was studied in Ref.~\onlinecite{Glazman}
where it was shown that the resistance between two droplets behaves
as \begin{align}
\frac{{\cal R}(B)}{{\cal R}(0)} & =e^{(B/B_{0})^{2}}\frac{1}{\cosh^{2}(B/B_{1})},\label{eq:raikhglazman}\end{align}
 where $B_{0}$ and $B_{1}$ have the following approximate expressions
in terms of our droplet parameters:\begin{align}
B_{0} & \sim\phi_{0}/(\pi y_{0}l_{ip}),\nonumber \\
B_{1} & \sim2\phi_{0}/(\pi l_{ip}^{2}),\label{eq:glazmanB0B1}\end{align}
 and we note that $l_{ip}=2\sqrt{R_{c}R_{p}}$, and we estimate that
the spread of the wavefunction under the barrier in the direction
perpendicular to the tunneling is $y_{0}\sim\sqrt{\xi r}.$ Both $B_{0}$
and $B_{1}$ are determined by the field for which a flux quantum
is enclosed within an area that is of significance in the droplet
picture. $B_{0}$ is the field that encloses a flux quantum in an
area of the order of the area enclosed by the electron wavefunction
tunneling under the barrier. $B_{1}$ is the field that encloses a
flux quantum in an area of the order of the {}``catchment region''.
{[}See Fig. \ref{fig:droplets}]. Physically, $B_{1}$ is the field
below which interference effects are significant. If $B_{0}>B_{1},$
then at small fields, the magnetoresistance will be negative as in
Eq.~(\ref{eq:negmagnres}), with $B_{c}^{-2}=B_{1}^{-2}-B_{0}^{-2}.$
Table \ref{tab:B0B1} lists $B_{0}$ and $B_{1}$ for various electron
densities.

\begin{table}
\begin{tabular}{|c|c|c|c|c|c|}
\hline 
$n_{e}$(cm$^{-2}$)&
$B_{0}$(T)&
$B_{1}$(T)&
$D$(nm)&
 \begin{tabular}{c}
$\alpha\,({\rm T}^{-2}),$\tabularnewline
$D=(3l_{ip}^{2}r/4)^{1/3}$\tabularnewline
\end{tabular}&
\begin{tabular}{c}
$\alpha\,({\rm T}^{-2}),$\tabularnewline
$D=r$\tabularnewline
\end{tabular}\tabularnewline
\hline
\hline 
$10^{11}$&
1.45&
0.22&
33&
0.48&
$9.6\times 10^{-4}$\tabularnewline
\hline 
$5\times10^{10}$&
0.42&
0.12&
95&
5.7&
0.84\tabularnewline
\hline 
$2\times10^{10}$&
0.19&
0.05&
180&
29&
13\tabularnewline
\hline
$10^{10}$&
0.10&
0.024&
271&
94&
62\tabularnewline
\hline
\end{tabular}

\caption{Magnetic fields $B_{0}$ and $B_{1}$ as defined in Eq. (\ref{eq:glazmanB0B1})
for different electron densities. These fields determine the crossover
between negative and positive magnetoresistance. Also shown are the
inter-droplet tunneling distance $D=(3l_{ip}^{2}/4)^{1/3}$ and $\alpha,$
both of which appear in magnetoresistance expressions in Sec. \ref{sec:Shklovskii}.
$D=(3l_{ip}^{2}r/4)^{1/3}$ is the tunneling distance obtained by
reconciling Eqs.~(\ref{eq:raikhglazman}) and (\ref{eq:glazmanB0B1})
with Eq.~(\ref{eq:shklovskiiRho1}). Values of $\alpha$ are also
shown for comparison by assumimg the tunneling distance $D$ is equal
to the distance $r$ between the surfaces of neighboring droplets. }

\label{tab:B0B1}
\end{table}
 For $n_{e}=5\times10^{10}\,{\rm cm}^{-2}$ we can infer from Table
\ref{tab:B0B1} that $B_{c}\simeq0.12\,{\rm T},$ which is close to the
experimentally observed $B_{c}\simeq0.15$ T.\cite{Baenninger1,Baenninger2} 
We note that recent
experiments on a quantum dot lattice\cite{Dorn} also displayed negative
magnetoresistance that appeared to be well explained by the Glazman
and Raikh approach.

The positive magnetoresistance in Eq.~(\ref{eq:raikhglazman}) arises
from the shrinking of the localization length in a strong magnetic
field. This has been extensively studied in the contest of dirty semiconductors.
In dirty semiconductors, the $e^{(B/B_{0})^{2}}$ dependence of magnetoresistance
is predicted to cross over to a $e^{B/B_{2}}$ dependence at a high
enough field. This is seen in the experiments we are discussing too,
while Eq.~(\ref{eq:raikhglazman}) makes no such prediction. The
high field data are better described by Shklovskii's expressions (see
Sec.~\ref{sec:Shklovskii}) for magnetoresistance for tunneling in
dirty semiconductors.\cite{Shklovskii1,Shklovskii2} This is the subject
of the following section.

\subsection{Larger fields: positive magnetoresistance}

\label{sec:Shklovskii}

The important length scale for magnetotransport in the regime of magnetic
fields where interference effects are negligible, is the length scale
of the {}``unit cell'' of the droplet array. What we mean is the
regime $l_{ip}>2l_{0}$, where $l_{ip}=2\sqrt{R_{c}R_{p}}$
is the separation between the centers of neighboring droplets, and
$l_{0}=\sqrt{\hbar/eB}$ is the magnetic length.

Now, it is well known from Refs.~\onlinecite{Shklovskii1} and \onlinecite{Shklovskii2}
that once the magnetic field is large enough, $B>B_{1},$ so that
quantum interference effects are negligible in comparison with the
positive magnetoresistance arising from the reduction of the localization
length, the magnetoresistance takes the form

\begin{equation}
\frac{{\cal R}(B)}{{\cal R}(0)}\sim e^{\alpha B^{2}},\label{eq:shklovskiiRho1}\end{equation}
 in the field range \begin{equation}
\frac{2\phi_{0}}{\pi l_{ip}^{2}}<B\ll\frac{\hbar}{e\xi D}.\end{equation}
 In Eq.~(\ref{eq:shklovskiiRho1}), with $D$ the typical tunneling
distance as before, $\alpha$ is given by

\begin{eqnarray}
\alpha & = & \frac{D^{3}\xi e^{2}}{3\hbar^{2}}.\label{eq:shklovskiiRho2}\end{eqnarray}
 If we take $\alpha$ to be determined from Eq.~(\ref{eq:raikhglazman}),
then we can identify $\alpha=B_{0}^{-2},$ and the tunneling distance
\begin{align}
D & =[3l_{ip}^{2}r/4]^{1/3}.\label{eq:D}\end{align}
 The upper limit for $B$ for the validity of Eq.~(\ref{eq:shklovskiiRho1}),
$\hbar/(e\xi D),$ is about $0.8\,{\rm T}$ for $n_{e}=5\times10^{10}{\rm cm}^{-2}.$
The quantitative estimate for $\alpha$ is very sensitive to $r$
and $l_{ip}$ which limits its reliability. Using the extrapolation
of Eq.~(\ref{eq:raikhglazman}), one should observe \begin{align}
\alpha & =\frac{A}{n_{e}^{\frac{3}{2}}}\left(1-\frac{R_{p}^{\frac{1}{2}}}{R_{c}^{\frac{1}{2}}}\right)\sim\frac{A}{n_{e}^{3/2}},\label{eq:shklovsviiExpt}\end{align}
 where $A=2n_{d}^{3/4}R_{p}^{3/2}\xi e^{2}/(\pi^{3/4}\hbar^{2}).$
A similar calculation, taking Shklovskii's expression for $\alpha$,
with $D=r$, also yields $\alpha\simeq A^{\prime}/n_{e}^{\frac{3}{2}}$,
where $A^{\prime}$ and $A$ differ by a factor of order unity but
have the same dependence on $n_{d}$, $R_{p}$, and $\xi$. In both
scenarios, the tunneling distance $D$ increases as $1/n_{e}^{1/2}$
as the electron density decreases, indicating that the $n_{e}$ dependence
of $\alpha$ is robust; additionally, $\alpha$ is seen to be independent
of temperature. This is exactly what was observed in Ref.~\onlinecite{Baenninger2},
although in that case the $n_{e}^{-\frac{3}{2}}$ dependence of $\alpha$
was ascribed to a Wigner crystal (or a charge density wave) where
the density of localized states is expected to be equal to the density
of electrons. We note that for the experimental parameters in Ref.~\onlinecite{Baenninger1},
there is quantitative agreement between our prediction for $\alpha$
and the observed behavior. In particular, considering $D$ as given
by Eq.(\ref{eq:D}), we have for $n_{e}\simeq10^{11}\,{\rm cm^{-2}}$,
$\alpha\simeq1.5\times10^{22}\,{\rm m^{-3} T^{-2}}/n_{e}^{\frac{3}{2}}$,
and if $n_{e}\simeq5\times10^{10}\,{\rm cm^{-2}}$, $\alpha\simeq6.4\times10^{22}\,{\rm m^{-3} T^{-2}}/n_{e}^{\frac{3}{2}}$.
These values are about an order of magnitude of the values quoted
in Ref.~\onlinecite{Baenninger1}, however, we note as in Sec.~\ref{sec:droplet}
that one of the main impediments to greater quantitative comparison
with experiment is an inability to get a better estimate for $R_{p}$,
which in turn affects our estimate of $\xi$. If one assumes that
the tunneling distance is equal to the distance between the neighboring
droplet surfaces, $D=r,$ then one obtains for 
$n_{e}\simeq5\times10^{10}\,{\rm cm^{-2}}$ that $\alpha\simeq9.4\times10^{21}\,{\rm m^{-3} T^{-2}}/n_{e}^{\frac{3}{2}}$, which is 
closer to experiment.

In our droplets picture such a behavior of $D$ arises due to the
$n_{e}$ dependence of the typical droplet spacing, and the $n_{e}$
dependence of $\alpha$ in Eq.~(\ref{eq:shklovsviiExpt}) is one
of our main results.

The expression for the magnetoresistance, Eq.~(\ref{eq:shklovskiiRho1}),
is valid when $D\ll l_{0}^{2}/\xi$, and for higher magnetic fields
where $D\gg l_{0}^{2}/\xi,$ (or equivalently, $B\gg\frac{\hbar}{e\xi D}$)
it crosses over to another regime,\cite{Shklovskii1,Shklovskii2}

\begin{eqnarray}
\frac{{\cal R}(B)}{{\cal R}(0)} & \sim e^{B/B_{2}},\label{eq:shklobskiiLargeB}\end{eqnarray}
 where \begin{align}
B_{2} & =\frac{\hbar}{eD^{2}}.\label{eq:B2}\end{align}
 In this regime of magnetic field, the result of Ref.~\onlinecite{Glazman}
presented in Eq.~(\ref{eq:raikhglazman}) is no longer valid. Note
also that the magnetoresistance expression, Eq.~(\ref{eq:shklobskiiLargeB}),
rather than Eq.~(\ref{eq:shklovskiiRho1}) or Eq.~(\ref{eq:raikhglazman}),
describes the correct behavior as the electron density is decreased.
Our estimate for $B_{2}$ is $0.07\,{\rm T}$ when $D\simeq95$
nm. In the experiments,\cite{Baenninger2} a crossover from a quadratic
to a linear magnetic field dependence (in the exponent) has been observed,
and it is in qualitative agreement with the expected crossover $B\sim\frac{\hbar}{e\xi D}\sim0.8$
T, for the parameters mentioned above. Again, we are able to predict
the dependence of this field on $n_{e}$; at low electron density,
\[
B_{2}\propto\frac{\hbar n_{e}}{en_{d}^{\frac{1}{2}}R_{p}}.\]
 This implies that the resistance should increase more quickly as
a function of magnetic field in samples with lower $n_{e}$, which
is observed in experiment.\cite{Baenninger2}

The picture we have discussed in this section, is one in which the
transport at increasing magnetic fields is dominated by barriers between
droplets. As we have seen, the barrier transparency drops exponentially
as the magnetic field is increased. This suggests that for large enough
magnetic fields, individual droplets will be relatively well isolated
from each other, and that one can think of the droplets as an irregular
array of quantum dots. This has further implications for transport
which we will discuss in future work.\cite{VMprep}

\section{Discussion}

\label{sec:conc}

The picture of electron droplets we have discussed here provides an
appropriate framework for describing the phenomenology of the recent
experiments in Ref.~\onlinecite{Baenninger2}. As we outline in
Sec.~\ref{sec:Shklovskii}, we are able to use the droplet picture
to make both qualitative and quantitative comparisons with experiment.
{[}We note that some of the quantitative agreement may be fortuitous,
however the dependence on 2DEG parameters should not be.] Importantly,
we are able to reproduce the dependence of the magnetoresistance on
the electron density. This strong agreement with experiment, in conjunction
with our analysis of the electronic environment in the 2DEG, strongly
implies that a picture of electronic droplets is more suitable to
describe these, and similar experiments than a charge density wave
scenario.

One point that we did not discuss quantitatively, was how the non-linear
screening is affected by the presence of a magnetic field. There has
been some discussions of screening 2D electrons in a disordered potential
in a magnetic field,\cite{Efros1,Efros2,EPB} but this has tended
to focus on the regime in which disorder is not too strong. We regard
this as a very interesting problem that merits further investigation;
however, we note that experiment appears to provide some of the solution
(hence our neglect of the issue here). Measurements of localized states
in the quantum Hall regime\cite{Ilani2} that support a dot-like picture
at low densities, find that the local electronic compressibility is
essentially independent of the magnetic field, supporting our assumptions
here.

 In our analysis we have assumed that the dopants are completely ionized. The degree of ionization is determined by the internal field at the GaAs-AlGaAs junction, the external gate voltage $V_g$, and the temperature $T_0$ at which the electron distribution over the dopants is frozen.
     Suppose the conditions are such that the donors are not completely ionized. In particular, consuder that the gate voltage $V_g$ is zero and that the electron distribution over the dopants is frozen at a nonequilibrium temperature $T_0$ which is less than the RMS potential fluctuation in the $\delta-$layer.
This case has been studied, for example, in Ref.~\onlinecite{Pikus}, where the
authors calculate the dopant atoms' correlator,
\begin{align}
D(\mathbf{r} - \mathbf{r'}) & = \langle n(\mathbf{r})n(\mathbf{r'})\rangle - \langle n\rangle^{2},
\label{eq:Dr}
\end{align}
using a path integral approach. The probability of a fluctuation
$c(\mathbf{r})=n(\mathbf{r})-\langle n \rangle$ is proportional
to $\exp(-\Phi[c])$, where
\begin{align}
\Phi[c] & = \frac{1}{2n_{d}}\int d\mathbf{r} c^{2}(\mathbf{r})
              + \frac{1}{k_{B}T_{0}}\int d\mathbf{r}\,d\mathbf{r'}
              c(\mathbf{r})G(\mathbf{r}-\mathbf{r'})c(\mathbf{r'});
\label{eq:Phi}
\end{align}
$G(\mathbf{r}-\mathbf{r'})$ is the interaction energy of two electrons
in the $\delta-$layer:
\begin{align}
G(\mathbf{r}-\mathbf{r'}) & = \frac{e^{2}}{4\pi\epsilon_{0}\kappa}
\left[\frac{1}{|\mathbf{r}-\mathbf{r'}|}
-\frac{1}{|(\mathbf{r}-\mathbf{r'})^{2}+4(d-\lambda)^{2}|}\right].
\label{eq:Gr}
\end{align}
   The dopant atoms' correlator is then given by
\begin{align}
D(\mathbf{r}-\mathbf{r'}) & =
\left[ \int Dc \,c(\mathbf{r})c(\mathbf{r'})\,e^{-\Phi[c]}\right]/
      \left[\int Dc \,e^{-\Phi[c]}\right]
\label{eq:Dpathint} 
\end{align}
Its Fourier transform $D(\mathbf{q})$ can be shown to be
\begin{align}
D(q) = \frac{n_{d}q}{q+q_{0}[1-\exp[-2q(d-\lambda)]]},
\label{eq:Dq}
\end{align}
where $q_{0}=n_{d}e^{2}/(2\epsilon_{0}\kappa k_{B}T_{0})$ is the
reciprocal Debye radius. In Ref.~\onlinecite{Pikus}, $T_0=100$K; for
this value we may assume that $q_{0} \gg q$. In the absence of 
dopant correlation, we have $D(q)=n_{d}$ as in Eq.(\ref{eq:gaussian}).
For $q(d-\lambda) \ll 1$, the correlator $D(q)$ approaches a constant value,
\begin{align}
D(q) & \approx n_{0} = \frac{\epsilon_{0}\kappa k_{B}T_{0}}{e^{2}(d-\lambda)}.
\label{eq:Dqeff}
\end{align} 
We estimate $n_{0}=2.5\times 10^{9}{\rm{cm}}^{-2}= 2\times 10^{-3} n_{d}$. 
Note that the correlator in Eq.(\ref{eq:Dqeff}) has the same form as 
the uncorrelated case, Eq.(\ref{eq:gaussian}), except that the dopant 
density $n_{d}$ is replaced by a smaller effective dopant density $n_{0}$. 
This implies that the potential fluctuations in the 2D electron layer (see Eq.(eq:meansqpotlimits)) are reduced, and the characteristic length scale beyond which the donor charges are uncorrelated is much larger. The $R_c$ corresponding to the reduced density is smaller than the original value by a factor of 20. Thus a much smaller electron density in the 2D layer suffices to screen the potential fluctuations from correlated dopants. The puddle size $R_p$ is relatively less affected by dopant correlation. Lower values of $R_c$ will take the system closer to metallicity since the puddles will merge once $R_c$ becomes smaller than $R_p$.
In the experiments we have studied, the observations are better explained 
by assuming a complete ionization of the donors.

   We also find it worthwhile to say a few words on the effect of sample width 
on the puddles scenario. This has been studied numerically, for example, in Ref.~\onlinecite{Nixon}. The conclusion is that a reduction of the sample 
width diminishes the effectiveness of screening of the potential fluctuations due to the dopant atoms, and the resulting enhanced potential fluctuations make 
puddle formation easier. In fact, the inter-puddle
separation will increase. For studying ballistic transport in quantum wires, one should must work in a regime where the potential fluctuations are lower. This can be made possible by working with, say, gate voltages $V_{g}$ where the dopants are incompletely ionized.

It is interesting to view our results in the broader context of other
physical phenomena observed in systems with nanoscale electronic disorder.
This includes other effects observed in low density delta-doped 2DEGs,
and in systems such as the underdoped cuprates.\cite{Davis} The picture
of droplets of charge that merge as the doping is increased may also
have relevance to the 2D metal-insulator transition.\cite{2DMIT,Meir,DasSarma}
It is interesting to note that experiments on the insulating side
of this transition have noted small jumps in the chemical potential
as a function of $n_{e}$,\cite{Ilani1} and we speculate that these
small jumps may be relevant to the small amplitude oscillatory behavior
of $\alpha$ as a function of $n_{e}$ observed in Refs.~\onlinecite{Baenninger1} and \onlinecite{Baenninger2}.
We note that our discussions here also lead to a plausible explanation
of recent observations of a zero-bias anomaly in insulating 2DEG samples
at low magnetic fields.\cite{Ghosh3,Ghosh4} This zero-bias anomaly
might be due to exchange interactions between electrons on two closely
coupled droplets having a Kondo effect similar to that observed in
double quantum dot systems.\cite{DQD} In the limit of higher density
and larger magnetic fields, our picture should evolve continuously
into existing droplet-based transport theory for the quantum Hall
regime.\cite{Nigel,Shimshoni} It would be interesting to try to connect
our work with this scenario. Finally, our analysis of strong disorder
in delta-doped heterostructures could perhaps also be extended to
a recent interesting proposal for creating strongly correlated electron
systems by modulation doping near a heterojunction of two Mott insulators.\cite{Lee_MacDonald}

\begin{acknowledgments}
The authors thank David Khmelnitskii, and Ben Simons for illuminating
discussions, and in particular thank Nigel Cooper for suggesting the
idea of charge droplets in the Ghosh experiments. We especially thank
Arindam Ghosh and Matthias Baenninger for many interesting discussions,
for critically reading this manuscript and for sharing their unpublished
data with us. V.T. acknowledges the support of Trinity College, Cambridge, and TIFR, Mumbai, and M.P.K. acknowledges support from NSERC. 
\end{acknowledgments}


\begin{thebibliography}{10}
\bibitem{Tsui} D. C. Tsui, H. L. Stormer, and A.C. Gossard, Phys. Rev.
Lett. \textbf{48}, 1559 (1982).

\bibitem{Yacoby1} S. Ilani, A. Yacoby, D. Mahalu, and H. Shtrikman,
Science \textbf{292}, 1354 (2001).

\bibitem{Yacoby2} J. Martin, S. Ilani, B. Verdene, J. Smet, V. Umansky,
D. Mahalu, D. Schuh, G. Abstreiter, and A. Yacoby, Science \textbf{305},
980 (2004).

\bibitem{Ilani2} S. Ilani, J. Martin, E. Teitelbaum, J. H. Smet,
D. Mahalu, V. Umansky, and A. Yacoby, Nature \textbf{427}, 328 (2004).

\bibitem{Wiebe} J. Wiebe, C. Meyer, J. Klijn, M. Morgenstern, and
R. Wiesendanger, Phys. Rev. B \textbf{68}, 041402(R) (2003).

\bibitem{Davis} S. H. Pan, J. P. O'Neal, R. L. Badzey, C. Chamon,
H. Ding, J. R. Engelbrecht, Z. Wang, H. Eisaki, S. Uchida, A. K. Gupta,
K.-W. Ng, E. W. Hudson, K. M. Lang, and J. C. Davis, Nature \textbf{413},
282 (2001); K. M. Lang, V. Madhavan, J. E. Hoffman, E. W. Hudson,
H. Eisaki, S. Uchida, and J. C. Davis, Nature \textbf{415}, 412 (2002);
K. McElroy, J. Lee, J. A. Slezak, D.-H. Lee, H. Eisaki, S. Uchida,
and J. C. Davis, Science \textbf{309}, 1048 (2005).

\bibitem{Various} S. Nakatsuji, V. Dobrosavljevic, D. Tanaskovic,
M. Minakata, H. Fukazawa, and Y. Maeno, Phys. Rev. Lett. \textbf{93},
146401 (2004); T. Park, Z. Nussinov, K. R. A. Hazzard, V. A. Sidorov,
A. V. Balatsky, J. L. Sarrao, S. W. Cheong, M. F. Hundley, J.-S. Lee,
Q. X. Jia, and J. D. Thompson, Phys. Rev. Lett. \textbf{94}, 017002 (2005).

\bibitem{Manganites} E. Dagotto \textit{et al.}, Phys. Reports \textbf{344},
1 (2001); N. Mathur and P. B. Littlewood, Phys. Today \textbf{56},
25 (2003); V. B. Shenoy, T. Gupta, H. R. Krishnamurthy, and T. V.
Ramakrishnan, cond-mat/0606660.

\bibitem{Baenninger1} M. Baenninger, A. Ghosh, M. Pepper, H. E. Beere,
I. Farrer, P. Atkinson, and D. A. Ritchie, Phys. Rev. B \textbf{72},
241311(R) (2005).

\bibitem{Baenninger2} M. Baenninger \textit{et al.} unpublished.

\bibitem{Ghosh2} A. Ghosh, M. Pepper, H. E. Beere, and D. A. Ritchie,
Phys. Rev. B \textbf{70}, 233309 (2004).

\bibitem{Ghosh3} A. Ghosh, C. J. B. Ford, M. Pepper, H. E. Beere,
and D. A. Ritchie , Phys. Rev. Lett. \textbf{92}, 116601 (2004).

\bibitem{Ghosh1} A. Ghosh, M. Pepper, H. E. Beere, and D. A. Ritchie,
J. Phys. Cond. Mat. \textbf{16}, 3623 (2004).

\bibitem{Suris} V. A. Gergel' and R. A. Suris, Zh. Eksp. Teor. Fiz.
\textbf{75}, 191 (1978). {[}Sov. Phys. JETP \textbf{48}, 95 (1978)].


\bibitem{Nixon} J. A. Nixon and J. H. Davies, Phys. Rev. B \textbf{41},
7929 (1990).

\bibitem{Shi} J. Shi and X. C. Xie, Phys. Rev. Lett. \textbf{88},
086401 (2002).

\bibitem{Fogler} M. M. Fogler, Phys. Rev. B \textbf{69}, 121409(R) (2004);
\textit{ibid}, \textbf{69}, 245321 (2004); \textit{ibid} \textbf{70},
129902 (2004).

\bibitem{Efros1} A. L. Efros, Solid State Commun. \textbf{67}, 1019
(1988).

\bibitem{Efros2} A. L. Efros, Solid State Commun. \textbf{70}, 253
(1989).

\bibitem{EPB} A. L. Efros, F. G. Pikus, and V. G. Burnett, Phys.
Rev. B \textbf{47}, 2233 (1993).

\bibitem{Glazman} M. E. Raikh and L. I. Glazman, Phys. Rev. Lett.
\textbf{75}, 128 (1995).

\bibitem{Shklovskii1} B. I. Shklovskii and A. L. Efros, Zh. Eksp.
Teor. Fiz. \textbf{84}, 811 (1983). {[}Sov. Phys. JETP \textbf{57},
470 (1983)].

\bibitem{Shklovskii2} B. I. Shklovskii, Fiz. Tekh. Poluprovdn \textbf{17},
2055 (1983). {[}Sov. Phys. Semicond. \textbf{17}, 1311 (1983)].

\bibitem{Pikus} F. G. Pikus and A. L. Efros, Zh. Eksp. Teor. Fiz.
\textbf{96}, 985 (1989). {[}Sov. Phys. JETP \textbf{69}, 558 (1989)].

\bibitem{nguyen1}V. L. Nguen, B. Z. Spivak, and B. I. Shklovskii,
JETP Lett. \textbf{41}, 42 (1985); Sov. Phys. JETP \textbf{62}, 1021
(1985).

\bibitem{sivan}U. Sivan, O. Entin-Wohlman, and Y. Imry, Phys. Rev.
Lett. \textbf{60}, 1566 (1988); O. Entin-Wohlman, Y. Imry, and U.
Sivan, Phys. Rev. B \textbf{40}, 8342 (1989).

\bibitem{schirmacher}W. Schirmacher, Phys. Rev. B \textbf{41}, 2461
(1990).

\bibitem{Dorn} A. Dorn, T. Ihn, K. Ensslin, W. Wegscheider, and M.
Bichler, Phys. Rev. B \textbf{70}, 205306 (2004).

\bibitem{VMprep} V. Tripathi and M. P. Kennett, in preparation.

\bibitem{2DMIT} S. V. Kravchenko, G. V. Kravchenko, J. E. Furneaux,
V. M. Pudalov, and M. D'Iorio, Phys. Rev. B \textbf{50}, 8039 (1994);
S. V. Kravchenko, W. E. Mason, G. E. Bowker, J. E. Furneaux, V. M.
Pudalov, and M. D'Iorio, Phys. Rev. B \textbf{51}, 7038 (1995); S.
V. Kravchenko, D. Simonian, M. P. Sarachik, W. Mason, and J. E. Furneaux,
Phys. Rev. Lett. \textbf{77}, 4938 (1996); E. Abrahams, S. V. Kravchenko,
and M. P. Sarachik, Rev. Mod. Phys. \textbf{73}, 251 (2001).

\bibitem{Meir} Y. Meir, Phys. Rev. Lett. \textbf{83}, 3506 (1999);
Phys. Rev. B \textbf{61}, 16470 (2000); Phys. Rev. B \textbf{63},
073108 (2001).

\bibitem{DasSarma} S. DasSarma, M. P. Lilly, E. H. Hwang, L. N. Pfeiffer,
K. W. West, and J. L. Reno, Phys. Rev. Lett. \textbf{94}, 136401 (2005).

\bibitem{Ilani1} S. Ilani, A. Yacoby, D. Mahalu, and H. Shtrikman,
Phys. Rev. Lett. \textbf{84}, 3133 (2000).

\bibitem{Ghosh4} A. Ghosh, M. H. Wright, C. Siegert, M. Pepper, I.
Farrer, C. J. B. Ford, and D. A. Ritchie, Phys. Rev. Lett. \textbf{95},
066603 (2005).

\bibitem{DQD} D. Goldhaber-Gordon, H. Shtrikman, D. Mahalu, D. Abusch-Magder,
U. Meirav, and M. A. Kastner, Nature \textbf{391}, 156 (1998); S.
M. Cronenwett, T. H. Oosterkamp, and L. P. Kouwenhoven, Science \textbf{281},
540 (1998).

\bibitem{Nigel} N. R. Cooper and J. T. Chalker, Phys. Rev. B \textbf{48},
4530 (1993).

\bibitem{Shimshoni} E. Shimshoni and A. Auerbach, Phys. Rev. B \textbf{55},
9817 (1997).

\bibitem{Lee_MacDonald} W.-C. Lee and A. H. MacDonald, Phys. Rev. B \textbf{74}, 075106 (2006).
\end{thebibliography}
\end{document}